\documentstyle[preprint,aps]{revtex}
\begin{document}
\title{Effect of disorder on the Kondo behavior of thin Cu(Mn) films}

\author{T. M. Jacobs and N. Giordano}

\address{Department of Physics, Purdue University, 
West Lafayette, Indiana 47907-1396}

\maketitle
\begin{abstract}

We have studied the influence of disorder on
the Kondo effect in thin films of Cu(Mn), i.e., Cu doped with 
a small amount of Mn.  We find that the Kondo contribution to
the resistivity is suppressed when the elastic mean-free-path, $\lambda$, is
reduced.  While this is qualitatively similar to results found
previously by our group in a different material, our new experiments reveal 
in detail how this suppression
depends on both film thickness and $\lambda$.  These
results are compared with the theory of Martin, Wan, and Phillips.
While there is general qualitative agreement with this theory, there
appear to be some quantitative discrepancies.

\end{abstract}

\pacs{94PACS 72.10.Di, 72.15.Rn, 73.20.Fz}

\section{Introduction and Background}

The behavior of magnetic impurities in metals, and in particular
the Kondo effect \cite{kondo64}, has been of interest for several 
decades \cite{fischer70}.  Even so,
it is the object of much current work, as a number of important issues 
in this area remain unresolved.
One of these issues is the Kondo behavior in small structures;
that is, in thin films and narrow wires.  Several years ago,
experiments \cite{chen90,chen91,parpia92,blachly92,blachly93,blachly94,blachly94b,blachly95} 
revealed that the Kondo effect can
depend on system size.  The Kondo effect makes a contribution to the
resistivity which, at high temperatures, has the form
\begin{equation}
\Delta \rho _K = -B_K \ln ( T ) ~.
\label{kondo_log}
\end{equation}
The conventional Kondo effect leads to an increase in the resistivity
at low temperatures ($\Delta \rho _K > 0$), so 
the coefficient $B_K$ is positive.
The experimental studies noted above
concerned $\Delta \rho _K$ in thin films and narrow wires, and
found that $B_K$ becomes smaller when the system size is reduced;
that is, the Kondo effect is suppressed in small
systems \cite{ibm_caveat}.  These experiments also revealed that
$B_K$ depends on the level of disorder, with its value decreasing
as the elastic mean-free-path, $\lambda$, is made smaller.  
Quite recently, a similar size dependent
suppression of the Kondo effect was observed in the
thermopower \cite{haesen98}.

Qualitatively, the Kondo effect is due to the screening of a magnetic
impurity by the conduction electrons \cite{kondo64,fischer70},
and it is the interaction responsible for this screening which
leads to (\ref{kondo_log}).  It was initially suggested \cite{chen91}
that the observed size dependence of $B_K$ might be related to the spatial
extent of the conduction electron screening cloud.
However, subsequent experiments \cite{blachly94,blachly95}
showed that this explanation cannot be correct,
as the length scale associated with
the suppression does not vary with $T_K$ in the manner expected from
this general picture.  In addition, it has been argued on
theoretical grounds that screening cloud physics cannot explain the
observed size dependence of $\Delta \rho _K$ \cite{bergmann,affleck}.
While the experiments did not suggest the mechanism responsible for
the size dependence, they did reveal that $\Delta \rho _K$ can also
be suppressed by an increase in the level of disorder, i.e., a
reduction of the elastic mean-free-path, $\lambda$.

The observations that $\Delta \rho _K$ depends on the system
size and on $\lambda$ have recently been addressed by two theoretical
studies.  An explanation for the size 
dependence in the clean (large $\lambda$)
limit has been developed by Zawadowski and 
co-workers \cite{zawadowski95}.  According to
their picture, a combination of spin-orbit scattering and
scattering from a surface (both involving the conduction electrons)
gives rise to a uniaxial anisotropy at a magnetic impurity.  This anisotropy
has the form $D S_z ^2$, where $D$ is a function of material parameters
(such as the density of states) and distance from the surface.
The precise value $D$ is very difficult to estimate, so the theory cannot
be judged or tested based on its prediction of the 
precise value of $D$.  However,
we have recently tested this theory in another way.  
In the experiments mentioned above, the magnetic species had integer
spin; in most cases it was Fe, which is believed to be described by
an effective spin $S = 2$ \cite{fred}.  In this case the anisotropy
energy splits the Fe sublevels, leaving a singlet $S_z = 0$ at an energy
$D$ below the $S_z = \pm 1$ doublet.
When $D$ is sufficiently large compared to $k_B T$, only this singlet
ground state will be occupied, and the Fe will be nonmagnetic.  This
is the origin of the suppression of $B_K$; as a system is made smaller,
an increasing fraction of the local ``moments'' are near the surface, and
are thus rendered nonmagnetic by the splitting arising from this
anisotropy.
In recent work, we investigated
the behavior of a magnetic impurity with half-integer spin, Mn which
has $S = 5/2$.  In this case the ground state will always be a doublet,
and hence magnetic.
Hence, the theory predicts that in this case $B_K$ will not vanish
as the system size $\rightarrow 0$, and this was indeed observed in our
experiments with Cu(Mn) \cite{jacobs98}.

In this paper, we again consider the behavior of
Cu(Mn), but now we focus on the behavior of $B_K$ as a function
of disorder.
In the case of strong disorder,
Martin, Wan, and Phillips \cite{phillips} have shown theoretically that
there is an interplay between Kondo physics and the quantum interference
effects which are responsible for weak localization \cite{bergmann84}.
This interplay gives $B_K$ the form
\begin{equation}
B_K = B_K ^ 0 \left(1 - \frac{\alpha}{t \lambda^2 \tau_s}\right) ~,
\label{model}
\end{equation}
where $B_k ^0$ is the value found in bulk systems,
$t$ is the film thickness, $\lambda$ is the elastic mean free path, 
$\tau_s$ is the spin scattering time, and
$\alpha = 1.2 \hbar / \pi m k_{F}$ is a parameter which depends only
on Fermi surface properties and is not expected vary with
disorder.
This result was obtained with the assumption that
$\lambda < t < L_\phi$,
where $L_\phi$ is the electron phase coherence length (which will
be discussed further below) \cite{martin}.
Martin, Wan, and Phillips have shown that the prediction (\ref{model})
gives a good account of
previous experimental results from our group for Cu(Fe) films \cite{phillips}. 

In the present paper we present results for the behavior of
$B_K$ for Cu(Mn) as a function of $\lambda$ and $t$, along with
independent measurements of $\tau_s$ via weak localization experiments.
This has allowed us to test the prediction (\ref{model}) in detail; we will
see that it does not seem to provide a very good quantitative
account of our new results.

\section{Experimental Method}

Cu(Mn) has been studied a great deal
in connection with the Kondo effect in bulk alloys \cite{fischer70}, making 
it an ideal choice for
the present experiments.  From previous work, we know that
a Mn concentration, $c$, in the neighborhood of a few hundred
parts per million (ppm) should be low
enough to observe behavior in the dilute limit; i.e., 
$\Delta \rho _K / c$ should be independent of $c$ \cite{monod}.
In order to produce films with high disorder we employed DC sputtering.
To obtain Cu films with a small concentration
of Mn impurities, several small pieces of manganin wire
(approximately 2 mm in length and 0.5 mm diameter) were placed uniformly
on the surface of a pure (99.999\%) Cu sputtering target of diameter 5~cm.
Manganin wire has the composition 86\% Cu, 12\% Mn, and 2\% Ni and
thus allows us to sputter a small amount of Mn along with the Cu.
Using this method has two slight drawbacks. (1) the films will contain
a small amount of Ni. However, since Ni does not have a local moment
when placed in Cu, we do not expect its presence to affect our 
results \cite{ni_caveat}.
(2) It is difficult to know, with great
precision, the Mn concentration.  This uncertainty
is due to the fact that
the target does not sputter uniformly over long periods of
time, and the different materials have different sputtering rates,
etc.  We have dealt with this problem in two ways.  First,
we will make direct comparisons only between samples prepared in
the same sputtering session.  In a single such session we deposited
a series of films, with different thicknesses.  This should
produce films with the same Mn concentration, but with
different thicknesses as determined by the time they are exposed
to the sputtering beam, and different disorder as determined
by the Ar pressure.  Second, during each sputtering session
the first and last films that were prepared were designed to
have the same mean-free-path and thickness.
We found that they always exhibited the same behavior (to within the
uncertainties), which demonstrates that the Mn concentration was indeed
constant throughout the session.

Based on the amount of manganin wire placed
on the target and the relative sputtering rates of Cu and Mn,
we estimate the concentration to be in the neighborhood of, or somewhat less
than, 300~ppm.
With the added knowledge that the data show no sign
of magnetic ordering or coupling effects \cite{chris}, 
we are confident of being in the dilute impurity regime. 

In a sputtering session, a collection of substrates (glass) was
mounted on a rotating holder system in such a way that only one substrate
at a time was exposed to the sputtering beam. The sputtering took place
with the substrates nominally at room temperature, with
Ar pressures in the range 1.5 to 10 mTorr.
After depositing a film, the sample holder
was then rotated without breaking vacuum to expose another substrate, and
the Ar gas pressure was adjusted to change the level of
disorder in the next film, with higher pressures yielding
films with greater disorder (i.e., shorter $\lambda$). 
This process was performed to make typically
six films using this target setup.  As noted above,
the first and last films were made at the same Ar pressure as a check on
our procedure.

Immediately after removing a given batch of films from the vacuum 
chamber they were coated with photoresist, and patterned with optical
lithography and etching in dilute nitric acid, to produce strips of
width $\sim$ 150 $\rm \mu m$ and length 60 cm. 
Note that the photoresist was then allowed to remain on the Cu(Mn)
films, thus protecting them from oxidation.
Resistance was measured as a
function of temperature using a standard 4-wire DC method in a
$^4$He cryostat.
Magnetoresistance measurements were made using an AC bridge technique
with the reference resistor either at room temperature or at the
same temperature as the sample.
    
The Kondo temperature of Cu(Mn) is not known precisely.
Previous work has established only that it is below
$\sim 10^{-2} ~ \rm K$.  This well below the range studied here,
so we will always be in the regime where the high
temperature limit (\ref{kondo_log}) is applicable. 

\section{Results and Discussion}

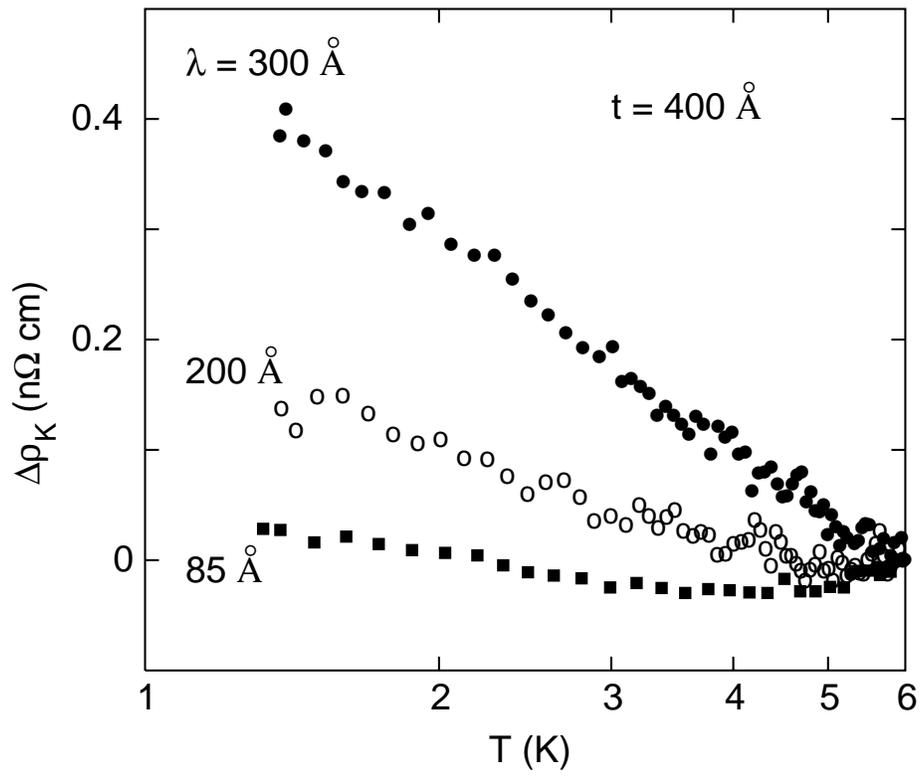
\begin{figure}
\caption{
Temperature dependence of the resistivity of several $400 ~ \AA$
thick films.  The mean-free-paths, $\lambda$, are indicated.  Note that
here we have plotted the {\em change} of the resistivity, as discussed
in the text, and the magnetic field here was zero.
\label{rho_T}
}
\end{figure}

Some typical results for the resistivity as a function of
temperature are shown in Fig.~\ref{rho_T}.
Here we have plotted just the {\em change} of the resistivity with
temperature, with the zero of $\Delta \rho$
chosen at a convenient temperature (here
near 6~K).  Below about 4~K it is seen that $\Delta \rho$
varies approximately logarithmically with $T$, as expected
from (\ref{kondo_log}).
At higher temperatures (not shown here)
the resistivity increases with increasing $T$, due to the usual effect
of electron-phonon scattering.  To avoid having to deal with this 
effect, we will restrict our attention to the behavior below
about 4~K, where electron-phonon scattering is negligible
compared to the Kondo contribution to $\Delta \rho$.

While the logarithmic variation seen in Fig.~\ref{rho_T} is
quite consistent with (\ref{kondo_log}), there are two 
other effects which can give rise to a similar temperature
dependence.  In two dimensions,
electron-electron interaction effects (EEI) \cite{lee85}
give rise to a logarithmic variation of
the sheet resistance, $R_ \Box$, which has the form
\begin{equation}
{{\Delta R _ \Box}\over {R _ \Box}} = - {{e^{2}} \over {2\pi^{2}\hbar}}
A_{ee} R_{\Box} \ln T ~,
\label{ee}
\end{equation}
where $A_{ee}$ is a screening factor which is typically near
unity in metal films \cite{lee85,monnier}.  Because of the dependence on
$R_\Box$, this contribution is much smaller than the Kondo
effect for large thicknesses.  So far as we know, it is believed that
the Kondo and EEI contributions are additive.  With this assumption,
we have subtracted the contribution calculated from (\ref{ee})
from our measurements (this subtraction was performed also for the
data in Fig.~\ref{rho_T}).  The screening factor $A_{ee}$ was
obtained from careful measurements
with pure Cu films (prepared under the sputtering conditions given above)
with resistivities and thicknesses similar to
the Kondo samples of interest.
This was accomplished by measuring the variation of the
resistivity (or equivalently, the sheet resistance, since the
thickness was known) with temperature.  By using pure Cu films, there
was no Kondo contribution, and by restricting the measurement to
low temperatures (below about 4~K) the electron-phonon contribution
discussed above was negligible.  This left only the EEI effect and
weak localization (WL), which we will discuss in detail in a moment.  The
WL effect was quenched (without affecting the EEI contribution)
by the application of a magnetic field of 15~kOe
perpendicular to the film plane.
From the results for several samples we found $A_{ee} = 1.0 \pm 0.1$,
a value which is quite in line with that reported previously for
similar films \cite{monnier}.  The uncertainty in $A_{ee}$
is a conservative estimate which encompasses all of our results for
Cu(Mn) films over a wide range of thickness.
This value for $A_{ee}$
will be used below to subtract the EEI contribution from the total
temperature variation found for our Cu(Mn) films.

Another contributor to our measured temperature dependence
is weak localization (WL).  This is a quantum interference 
effect \cite{bergmann84} which makes a contribution of the form
\begin{equation}
{{\Delta R_\Box} \over {R_\Box}} =
{{e^{2}} \over {4 \pi^{2}\hbar}} R_{\Box} \ln L_{\phi}(T) ~,
\label{wl}
\end{equation}
for our Cu and Cu(Mn) samples (which exhibit antilocalization
in small fields).
Here $L_ \phi$ is the electron phase coherence length.  
If, as happens to occur in many cases,
$L_\phi$ varies as a power of $T$, the WL contribution varies
logarithmically with temperature, with a magnitude similar to
that of the EEI effect.  However, in Cu(Mn) the phase coherence
length is limited by the effect of the process of spin 
scattering \cite{bergmann84},
in which the spin of a conduction electron is flipped through an
interaction with
the local moment (Mn).  This typically gives rise to phase coherence
length which varies only weakly, if at all, with temperature \cite{caveat1}.

In most cases below we are not interested in the WL contribution.
For our thickest samples, which have the smallest sheet resistances,
the WL effect is generally negligible (since the WL effect
(\ref{wl}) is proportional to $R_\Box$).  It becomes more important
for the thinnest films, but even in such cases it can be avoided
by performing the measurement in a magnetic field.  As noted above,
a large magnetic field applied perpendicular to the film quenches 
WL.  While WL can thus be easily avoided, it can also provide some
important information.  The theoretical prediction (\ref{model})
involves the spin scattering time, $\tau_s$.  When the phase coherence
length is dominated by spin scattering, it is given by
$L_\phi = \sqrt {D \tau_s}$, where $D = v_F \lambda / 3$ 
is the electron diffusion constant (here $v_F$ is the
Fermi velocity).
Measurement of the magnetoresistance in low fields, which is due
solely to WL, can be used to extract $L_\phi$ and hence also $\tau_s$.
We will make use of this fact below.

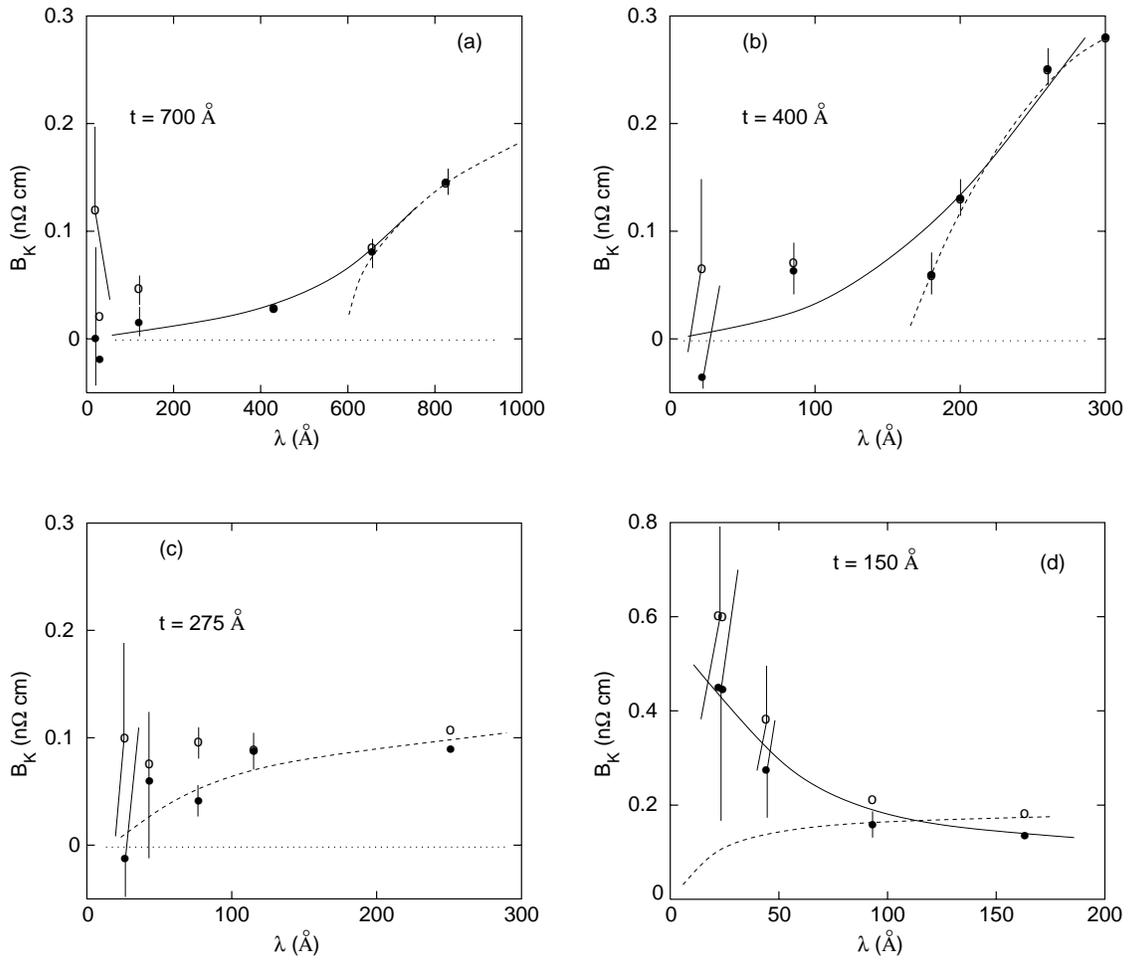
\begin{figure}
\caption{
Variation of the Kondo slope, as defined in (\protect{\ref{kondo_log}}),
as a function of $\lambda$ for different film thicknesses, $t$, as
indicated.
The open circles were obtained in zero magnetic field, while the
filled circles are data obtained with $H = 5 ~ \rm KOe$.  The
difference beween the two cases becomes smaller as $\lambda$
is increase, and for the thickest films, $t = 400$ and $700 ~ \rm \AA$,
the two results are in many cases indistinguishable for
large $\lambda$.
The solid and dashed lines are guides to the eye, and are discussed
in the text.
\label{thickness}
}
\end{figure}

Returning to Fig.~\ref{rho_T}, we see that varying the
level of disorder, i.e., $\lambda$,
has a large effect on the behavior.  This can also be seen
from Fig.~\ref{thickness}, which shows results for $B_K$ as a function
of $\lambda$, for different values of film thickness, $t$.
Each data point was obtained from a measurement of the resistivity, $\rho$,
as a function of $T$, like the ones shown in Fig.~\ref{rho_T}.  The
results for $\rho (T)$ were fit to a logarithmic form, and
the EEI effect was subtracted using the (previously) measured value of $A_{ee}$
discussed above.  
Note also that we give results for magnetic field
$H=0$, and for $H = 5~\rm kOe$ applied perpendicular to the plane of the
sample; the latter should be sufficient to quench WL.  Typical uncertainties
are shown; they are seen to become larger for the thinner films, for the
following reason.  As the film thickness decreases, the sheet resistance
increases, making EEI larger relative to the Kondo effect.  Our uncertainty
in $A_{ee}$ then leads to a larger uncertainty in $B_K$.
The solid curves in Fig.~\ref{thickness} are simply guides to the
eye drawn through the data for $H = 5~\rm kOe$ (the filled symbols).  
The dashed curves have the general function form predicted by
the theory (\ref{model}).

Before discussing the results in Fig.~\ref{thickness} we should note again
that all of the data in a given
plot were obtained from samples prepared in
a single sputtering session, and hence had the same Mn concentration.
However, this concentration varied somewhat from one session to the
next, so differences in the absolute scale of $B_K$ in the different
cases are likely due, completely or in large part, 
to variations in the concentration.
This should not affect either the general shapes of these curves as a function
of $\lambda$, or the variation with field,
and these are what we will rely on in our analysis below.

The thickest samples, with $t = 700$ and $400 ~ \rm \AA$, exhibit similar
behavior.  For large disorder, i.e., small $\lambda$,
$B_K$ appears to approach zero to within the uncertainties, except perhaps
for the smallest values of $\lambda$ where the uncertainties
(due to uncertainties in the EEI subtraction)
become very large.  As $\lambda$
is increased, $B_K$ also increases, with this increase becoming more rapid 
when $\lambda$ exceeds a ``threshold'' value.
This threshold is $\lambda \sim 500 ~ \rm \AA$ for the samples
with $t= 700 ~ \rm \AA$, and decreases to $\lambda \sim 150 ~ \rm \AA$ when
the thickness is reduced to $400 ~ \rm \AA$.  
It is especially noteworthy that the behavior for these two
values of the film thickness is affected very little by the
application of a magnetic field.  The only change
is a decrease in $B_K$ in the presence of a field; 
the magnitude of this decrease is very small
for the case of weak disorder (large $\lambda$) and becomes
larger as $\lambda$ is reduced.
Qualitatively and quantitatively, it would appear that the
only effect of a field is to quench weak localization.

The behavior for the thinner
samples is more difficult to determine, as the uncertainties are larger
(for the reasons discussed above).  The results for
the $275 ~ \rm \AA$ thick samples are consistent with
$B_K \rightarrow 0$ as $\lambda$ becomes small, or with
a $B_K$ which is independent of $\lambda$.
The uncertainties for the
$t = 150 ~ \rm \AA$ samples are also substantial.  Here again the
data are consistent (barely) with $B_K$ being independent of
disorder, although they
seem to prefer a value of $B_K$ which grows substantially
as $\lambda \rightarrow 0$.

Let us now compare these results to the theory (\ref{model}).
Taken at face value, (\ref{model}) would seem to predict that $B_K$
can become negative, i.e., a ``negative'' Kondo effect, when either
$\lambda$ is made sufficiently short, or the thickness $t$ is made
very small.  To within our uncertainties, we have no evidence for
a negative $B_K$ in these limits.  However, it seems quite plausible
that the expression (\ref{model}) cannot be extrapolated
to the regime where $\alpha / (t \lambda ^2 \tau_2) \sim 1$;
i.e., other higher order terms or contributions may then be important, etc.
If this is the case, then (\ref{model})
cannot be meaningfully extrapolated to the parameter
regime where it yields a negative $B_K$.  However, it may still
make sense to use it to estimate the ``threshold'' values
of $\lambda$ noted above.  That is, the values of $\lambda$
at which $B_K$ is observed in Fig.~\ref{thickness} to increase substantially
may be estimated from the condition 
\begin{equation}
{ {1} \over {t \lambda ^2}} \approx constant ~,
\label{estimate}
\end{equation}
which is obtained by simply setting (\ref{model}) to zero.
The condition (\ref{estimate}) suggests that the threshold value
of $\lambda$ should increase as the thickness is reduced.  However,
this is opposite the trend observed in Fig.~\ref{thickness},
as the thickness is reduced from 700 to 400$~\rm \AA$.
It does not appear that the trend found in our experiment can
be accounted for by (\ref{model}).

The effect of a field on $B_K$ is also noteworthy.  The disorder correction
to the bulk Kondo effect in (\ref{model}) arises from a Kondo
contribution to the spin scattering time which then affects WL.
According to this theoretical picture, application of a magnetic field should
not only quench the ``ordinary'' WL effect, but also destroy
the suppression of $B_K$.  This means that
a field should cause the Kondo contribution to {\em increase} with
respect to its value in zero field.  Such
behavior is contrary to what is observed
in our experiments, Fig.~\ref{thickness}, where $B_K$ is seen to
be either essentially constant, or decrease, in the presence of a field.

\subsection{Spin scattering rate}

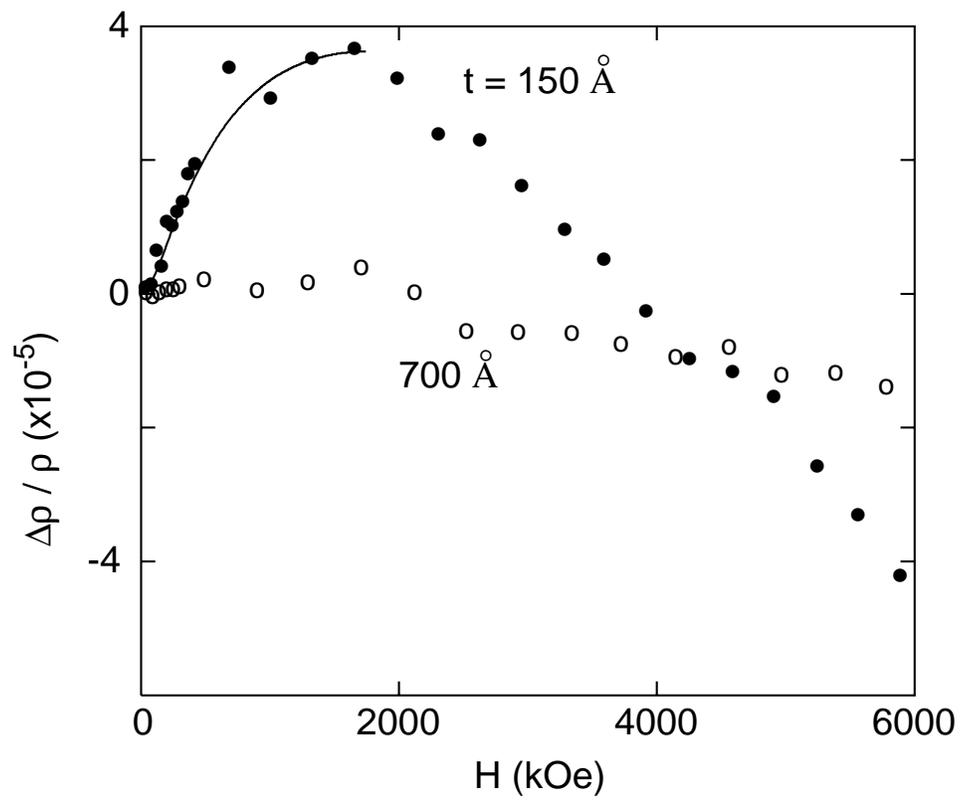
\begin{figure}
\caption{
Variation of the resistivity as a function of magnetic field of
two Cu(Mn) samples.  The temperature was 1.4~K, and the sample
thicknesses are indicated in the figure.  The solid line for the
$150 ~ \rm \AA$ sample shows a fit to the theorically expected
WL magnetoresistance.
\label{rho_H}
}
\end{figure}

The spin scattering time, $\tau_s$, plays a key role in the prediction
(\ref{model}).  We have therefore used measurements of the WL magnetoresistance
to obtain an independent measure of $\tau_s$.  Typical results
for the variation of the resistivity as a function of magnetic field
applied perpendicular to the film plane are shown in Fig.~\ref{rho_H}.
The analysis of such measurements are now standard, and are described
in detail elsewhere \cite{blachly95,beutler}.  A 
fit to the theoretical form for the WL magnetoresistance
yields the phase coherence length, $L_\phi$.
The results from the data in Fig.~\ref{rho_H} are $1500 ~ \rm \AA$
for the $150 ~ \rm \AA $ thick sample, and $1300 ~ \rm \AA$ for the
thicker sample, with uncertainties of approximately $100 ~ \rm \AA$
in both cases.  These values are in good accord with results
found previously for other samples in which spin scattering
was dominant \cite{blachly95,pennington}.  In addition, to within the
uncertainties just noted, the value of $L_\phi$ changed very little in
going to 4.2~K, which is also expected when spin scattering is dominant. 

To obtain $\tau_s$ from the phase coherence length, we must estimate
the value of the diffusion constant.  {\em If} we 
use nearly-free-electron theory to obtain the Fermi 
velocity \cite{kittel}, we find
$v_F = 1.6 \times 10 ^8 ~ \rm cm/s$, which for the samples in 
Fig.~\ref{rho_H} leads to $D \sim 15 ~ \rm cm^2/s$.  
Using this value together with the phase coherence
lengths found from the magnetoresistance gives 
$\tau_s \sim 1.5\times 10^{-11} ~ \rm s$.   

Let us now compare this with the theory by using (\ref{model}) to
calculate what value of $\tau_s$ would be needed to suppress $B_K$ to 
zero at the values of $t$ and $\lambda$ observed in the experiment,
Fig.~\ref{thickness}.  From the results for $t = 400 ~ \rm \AA$, we
estimate that $B_K \rightarrow 0$ at $\lambda \approx 150 ~ \rm \AA$.
Inserting these values into (\ref{model}) 
we find $\tau_s \sim 3 \times 10^{-9} ~ \rm s$ \cite{old_fit}.
This is approximately two orders of magnitude larger than
the measured $\tau _s$.  The measured
value of $\tau_s$ is derived from the phase
coherence length, and therefore depends
on the value employed for the diffusion constant, which contributes some 
uncertainty.  However, we do not believe that this uncertainty
amounts to two orders of magnitude.  The value of $\tau_s$ required
to make the theory (\ref{model}) compatible with our results is
difficult to reconcile with our direct measurement of the spin scattering
time.

\section{Conclusions}

We have found that the Kondo effect in thin Cu(Mn) films is suppressed
as the level of disorder is increased. This dependence on disorder
is similar to that found previously for Cu(Fe).
However, the present results are much more detailed, and are the
first to reveal the detailed dependence of $B_K$ on the film
thickness, elastic mean-free-path, and magnetic field.  The theory of Martin,
Wan, and Phillips provides only a qualitative account of our data.  There
are several, potentially serious,
quantitative discrepancies which remain unresolved, and which suggest
that the crucial physics for this problem is not yet accounted for.

\section*{Acknowledgements}
We thank A. Zawadowski, P. F. Muzikar, and especially
P. Phillips and I. Martin, for many enlightening, and patient, discussions.  
This work was supported 
by the National Science Foundation through grant DMR 95-31638.

\begin{references}
\bibitem{ibm_caveat}
We should note that one group has disputed
this result \cite{ibm94}, i.e., failed to observe a suppression of
$B_K$.  Sevaral authors have discussed how those experimental results
can be reconciled with the other experiments noted above.
In brief, it appears that a careful consideration of surface roughness
and also the surface to volume ratio, together with the 
theory \cite{zawadowski95,fomin} can account for the null result reported in
\cite{ibm94}.
\bibitem{ni_caveat}
In our studies of clean Cu(Mn) \cite{jacobs98}, 
we compared the behavior of samples
prepared using manganin wire (i.e., containing a small amount of Ni)
to that of Cu(Mn) without any Ni.  We found no discernible difference
in the behavior.
\bibitem{caveat1}
We assume here that the temperature is much higher than the Kondo 
temperature, which is here the case for Cu(Mn).
\bibitem{old_fit}
A similar value of $\tau _s$ was
inferred in \cite{phillips}, when the theory (\ref{model}) was
fit to data for Cu(Fe) from \cite{blachly95}.
\end{references}
\end{document}